# Micromagnetic simulations of absoption spectra

K. Rivkin[1] and J. B. Ketterson[1,2,3]

1. Department of Physics and Astronomy, Northwestern University
Evanston IL, 60201
2. Department of Electrical and Computer Engineering, Northwestern University
Evanston IL, 60201
3. Materials Research Center, Northwestern University
Evanston IL, 60201

**Abstract**

Further development of a previously introduced method for numerically simulating magnetic spin waves is presented. Together with significant improvements in speed, the method now allows one to calculate the energy absorbed by the various modes excited by a position- and time-dependent $\mathbf{H}_1$ field in a ferromagnetic body of arbitrary shape in the presence of a (uniform or non uniform) static $\mathbf{H}_0$ field as well as the internal exchange and anisotropy fields. The method is applied to the case of the single vortex state in a thin disc, a ring, and various square slabs, for which the absorption spectra are calculated and the most strongly excited resonance modes are identified.

PACS: 75.75.+a, 75.10.Hk, 75.30.Ds

**Introduction**

Over the last few decades a number of methods have been proposed that enable one to calculate the frequencies and spatial dependencies of the magnetostatic resonance modes of a ferromagnetic sample. While most models are macroscopic[i,ii], a microscopic model was proposed by Politi et. al.[iii] and subsequently used (for example by Nortmann et. al.[iv]) to study ferromagnetic resonances in various objects. In this model a macroscopic body is approximated by an array of magnetic dipoles, which are viewed as "block spins" representing an average over many crystallographic unit cells, rather than true microscopic spins. For disturbances with wavelengths large compared to a block-spin spacing we expect the method to provide a reasonable representation of the associated excitations.

Another interesting method has been developed by Kamenetskii.[v] However this method does not provide a description of how such objects interact with an applied position- and time-dependent $H_1$ magnetic field; i.e., which modes are excited, what is the amplitude of each excited mode, and how the absorption depends on the applied fields.

The approach generally chosen to answer the latter questions has been to implement a Runge-Kutta-like scheme to integrate the Landau-Lifshitz-Gilbert equation of motion for an ensemble of spins (as for example in the work of Jung et al. [vi]). This method has numerous advantages: it enables one to see the complete time evolution of the system starting with some initial conditions, and to utilize external fields having some specific time dependence as, e.g., in spin-echo experiments. It also has numerous disadvantages – it is relatively slow and only permits one to analyze a single frequency of the external driving field in a given run; any change requires the whole calculation be repeated.

Searching for a method that does not suffer these disadvantages has led to a revival of interest in using an eigenvalue-based formulation of the spin wave problem. Indeed, in the past year alone, together with attempts to use the ideas of Politi et al, two new methods have appeared – a method introduced by the authors,[vii] and a somewhat similar method by Grimsditch et. al.[viii] We note that that an analogous technique, referred to as the "discrete dipoles method" has been widely used in optics in recent years.

The most severe limitation of the eigenvalue-based approaches is that, unlike the Runge-Kutta approach, they cannot be used in the non-linear regime.

**2. Finding the resonance frequencies and modes.**

We begin with a brief review of our basic formalism (a more detailed discussion is given elsewhere[vii]). We recall the Landau-Lifshitz equation in the presence of a dissipative Gilbert term:

$$\frac{d\mathbf{m}_i}{dt} = -\gamma \mathbf{m}_i \times \mathbf{h}_i^{total} - \frac{\beta\gamma}{M_s} \mathbf{m}_i \times \left(\mathbf{m}_i \times \mathbf{h}_i^{total}\right); \tag{2.1}$$

here $\mathbf{m}_i$ is the magnetic moment of the $i^{th}$ spin, $\gamma$ is the gyromagnetic ratio, and $\beta$ is a parameter governing the dissipation; henceforth a "spin" or "discrete dipole" can imply a variety of objects, from discrete dipoles per se, to close-packed uniformly magnetized prisms. The strategy is to linearize the problem by writing both the applied fields and magnetic moments as the sum of a zeroth-order static part and a small first-order time-dependent perturbation:

$$\mathbf{m}_i = \mathbf{m}_i^{(0)} + \mathbf{m}_i^{(1)}(t) \tag{2.2a}$$

$$\mathbf{h}_i = \mathbf{h}_i^{(0)} + \mathbf{h}_i^{(1)}(t). \tag{2.2b}$$

With such a representation, the linearized form of Eq. (2.1) is given by:

$$-\frac{d\mathbf{m}_i^{(1)}}{dt} = \gamma\left[\mathbf{m}_i^{(0)} \times \mathbf{h}_i^{(1)} + \mathbf{m}_i^{(1)} \times \mathbf{h}_i^{(0)}\right] + \frac{\gamma\beta}{M_s} \mathbf{m}_i^{(0)} \times \left[\mathbf{m}_i^{(0)} \times \mathbf{h}_i^{(1)} + \mathbf{m}_i^{(1)} \times \mathbf{h}_i^{(0)}\right].$$

$$\tag{2.3}$$

We assume a solution of the form

$$\mathbf{m}_i^{(1)}(t) = \mathbf{m}_i^{(1)} e^{-i\omega t} \tag{2.4}$$

where

$$\omega = \omega' - i\beta\omega''; \tag{2.5}$$

here $\omega$ and $\mathbf{m}_i^{(1)}$ are, respectively, the eigenvalue and eigenvector of Eq. (2.3) and can be obtained by solving Eq. (2.3). For $\beta = 0$, corresponding to the dissipationless case, the eigenvalues $\omega$ will be real, provided the zeroth-order spin directions have been relaxed to their stable equilibrium directions. Note the orbits of the oscillations become progressively more circular as the external field increases.

Due to the fact that we use a complex form to represent the physical moments, which are real, any solution of Eq. (2.3) with a positive $\omega'$ and eigenvector $\mathbf{m}_i^{(1)}$, will be accompanied by a second solution with eigenfrequency $-\omega'$ and eigenvector $\mathbf{m}_i^{(1)*}$.

The fields entering Eq. (2.3) are given by the formulas:

$$\mathbf{h}_i^{(0)} = \mathbf{h}_i^{dipole} + \mathbf{h}_i^{exchange} + \mathbf{H}_0 = \sum_{j \neq i} \mathbf{m}_j^{(0)} \left( \nabla \nabla \left( \frac{1}{r_{ij}} \right) \right) + \mathbf{h}_i^{exchange} + \mathbf{H}_0 \quad (2.6)$$

and

$$\mathbf{h}_i^{(1)} = \mathbf{h}_i^{dipole} + \mathbf{h}_i^{exchange} = \sum_{j \neq i} \mathbf{m}_j^{(1)} \left( \nabla \nabla \left( \frac{1}{r_{ij}} \right) \right) + \mathbf{h}_i^{exchange}, \quad (2.7)$$

which account for both the dipole-dipole and exchange interactions; it is straight forward to add anisotropy energy terms. In the present paper we restrict the exchange interaction to nearest neighbors only; i.e.,

$$\mathbf{H}_{exchange} = J \sum_i^{NN} \mathbf{m}_i; \quad (2.8)$$

where $J = 2 \dfrac{A_{Gauss}}{M_s^2 a^2}$. (2.9)

We earlier used three complex numbers to describe the individual oscillations at each of the sites i.[vii] However in the linear regime, spin precession can be represented by an ellipse lying in a plane perpendicular to the equilibrium direction of the magnetic moment, a description of which requires only two complex numbers. One way to reduce the initial eigenvalue problem from 3N × 3N to 2N × 2N is to transform all magnetic moments to local coordinates, in which the z-axis coincides with the equilibrium direction of the magnetic moment at this point:

$$m_{i\alpha}^{(1)} = \sum_\beta A_{i\alpha\beta} m_{i\beta}^{(1)'} \quad (2.10)$$

where

$$A_{\alpha\beta} = \begin{pmatrix} \cos\beta\cdot\cos\alpha & -\sin\alpha & \sin\beta\cdot\cos\alpha \\ \cos\beta\cdot\sin\alpha & \cos\alpha & \sin\beta\cdot\sin\alpha \\ -\sin\beta & 0 & \cos\beta \end{pmatrix}$$

$$\cos\alpha = \frac{m_{ix}^{(0)}}{\sqrt{m_{ix}^{(0)2} + m_{iy}^{(0)2}}} \quad \sin\alpha = \frac{m_{iy}^{(0)}}{\sqrt{m_{ix}^{(0)2} + m_{iy}^{(0)2}}}$$

$$\cos\beta = \frac{m_{iz}^{(0)}}{\left|\mathbf{m}_i^{(0)}\right|} \quad \sin\beta = \frac{\sqrt{m_{ix}^{(0)2} + m_{iy}^{(0)2}}}{\left|\mathbf{m}_i^{(0)}\right|}$$

$$A_{\alpha\beta}^{-1} = A_{\alpha\beta}^{T}$$

(2.11)

Then Eq. (2.3) can be rewritten as :

$$i\omega m_{i\alpha}^{(1)} = \sum_{j,\beta} \gamma B_{ij\alpha\beta} m_{j\beta}^{(1)}$$

$$i\omega m_{i\alpha'}^{(1)\prime} = \sum_{j,\beta} \gamma B'_{ij\alpha'\beta'} m_{j\beta'}^{(1)\prime}$$

$$B'_{ij\alpha'\beta'} = A_{i\alpha'\alpha}^{-1} B_{ij\alpha\beta} A_{j\beta\beta'}$$

(2.12)

Another problem associated with the eigenvalue method is that every new spin (or cell) added to the system corresponds to a new mode. On the other hand, accurately describing the shape of a body requires a fine discretization (and hence a large number of spins). On the other hand, experimentally we can typically observe only the longer wavelength modes. The simplest way to restrict to these longer wavelength modes is to use "spatial averaging": assuming that certain cells oscillate in basically the same way (or that the resonance modes we are interested in are locally uniform), we can group these cells into bigger cells, later referred to as "macro-cells", with an averaged behavior assigned to each of the macro-cells.

We write Eq. (2.3) as:

$$i\omega m_{iji\alpha}^{(1)} = \sum_{j,\beta} \gamma B_{iji'j'\alpha\beta} m_{i'j''\beta}^{(1)}$$

(2.13)

where we identify each individual cell with two integers: i corresponds to the macro-cell, and j to the position inside the macro-cell. Then, using the assumption of nearly uniform behavior within each macro-cell,

$$\mathbf{m}_{ij}^{(1)} = \bar{\mathbf{m}}_i^{(1)}$$
$$\omega \bar{\mathbf{m}}_i^{(1)} = \sum_{i'} \frac{1}{N_i} B'_{ii'} \bar{\mathbf{m}}_{i'}^{(1)} \qquad (2.14)$$
$$B_{ii'} = \sum_j \sum_{j'} B_{iji'j'}$$

where $N_i$ is number of cells within macro-cell i.

We note that it is not necessary for the macro-cells to contain the same number of individual cells ("spins"). For example when we have a spherical object approximated by cubic cells, macro-cells on the boundaries will provide for a high discretization at the boundary, thereby providing a smoother boundary, which would look more "pixilated" if we simply used larger cells. This method proved to be extremely effective when studying modes in thin structures (slabs, discs etc.) that are uniform along the sample thickness; a comparison of the calculated absorption spectra showed that no significant errors are introduced by this method, and that it is far superior to simply using the larger cells. However, this method should be applied cautiously when the nature of modes is unknown; for example in a highly non-uniform vortex structure the method can potentially destroy the influence that the vortex core has on the modes, seriously changing the nature of the resonance spectrum.

## 3. Calculating the response to an applied time-dependent magnetic field.

An overly simplified, short introduction to this method have been presented by us in[ix]. After solving Eq. (2.3) for the eigenvectors $V_i^{(k)}$ and the corresponding eigenvalues $\omega^{(k)}$, we must then ask how the various modes of oscillation can be excited by applying a time-dependent external magnetic field. In the presence of this dynamic field, Eq. (2.3) can be written in a matrix form as:

$$-i\omega^{(k)}V_{i\alpha}^{(k)} = \gamma B'_{ij\alpha\beta}V_{j\beta}^{(k)}$$

$$\frac{dm_{i\alpha}^{(1)}}{dt} = \gamma B'_{ij\alpha\beta}m_{j\beta}^{(1)} + g_{i\alpha} \qquad (3.1)$$

$$g_{i\alpha} = \gamma\left(\mathbf{m}_i^{(0)} \times \mathbf{h}_i^{(rf)}\right)_\alpha = \gamma\varepsilon_{\alpha\beta\chi}m_{i\beta}^{(0)}h_{i\chi}^{(rf)}$$

here and in what follows we employ the summation convention, roman letters correspond to individual dipoles, greek letters correspond to coordinates.

We can transform this system to a simpler one

$$m_{i\alpha}^{(1)} = P_{i\alpha k}y_k \quad P_{i\alpha k} = V_{i\alpha}^{(k)} \quad P_{i\alpha k}^{-1} = V_{i\alpha}^{(k)*} \quad P^{-1}P = 1$$

$$y_k(t) = e^{\lambda^{(k)}t}c_k + e^{\lambda^{(k)}t}\int e^{-\lambda^{(k)}t}\sum_{i,\alpha} P_{i\alpha k}^{-1}g_{i\alpha}(t)dt. \qquad (3.2)$$

where $\lambda^{(k)} = -i\omega^{(k)}$ is an eigenvalue of a homogeneous system. We consider only the case where the applied r.f. field has a sinusoidal time-dependence:

$$g_{i\alpha}(t) = g_{i\alpha}(0)e^{-i\omega t}$$

$$y_k(t) = e^{\lambda^{(k)}t}c_k + e^{\lambda^{(k)}t}\frac{e^{-\lambda^{(k)}t}e^{-i\omega t}P_{i\alpha k}^{-1}g_{i\alpha}(0)}{-\lambda^{(k)} - i\omega} \qquad (3.3)$$

The steady-state solution is:

$$c_k = 0$$

$$m_{i\alpha}^{(1)} = P_{i\alpha k}y_k = -iV_{i\alpha}^{(k)}\frac{V_{l\beta}^{(k)*}\gamma\varepsilon_{\beta\sigma\chi}m_{l\sigma}^{(0)}h_{l\chi}^{(rf)}}{\omega^{(k)} - \omega}e^{-i\omega t} \qquad (3.4)$$

We can define a *magnetic susceptibility* by writing

$$m_{i\alpha}^{(1)}(t) = \chi_{ij\alpha\beta}h_{j\beta}^{(rf)}e^{-i\omega t} \qquad (3.5)$$

where

$$\chi_{ij\alpha\beta} = -i\gamma\sum_{\eta,\sigma,k} V_{i\alpha}^{(k)}\frac{V_{j\eta}^{(k)*}\varepsilon_{\eta\sigma\beta}m_{j\sigma}^{(0)}}{\omega^{(k)} - \omega}h_{j\beta}^{(rf)} \quad . \qquad (3.6)$$

Since physical magnetic fields are real, r.f. magnetic field must be represented as a sum

$$m_{i\alpha}^{(1)}(t) = V_{i\alpha k}\left(C_k^{(\omega)}e^{-i\omega t} + C_k^{(-\omega)}e^{i\omega t}\right) \qquad (3.7)$$

$$C_k^{(\omega)} = -i\gamma V_{i\alpha}^{(k)*}\varepsilon_{\alpha\beta\chi}m_{i\beta}^{(0)}\frac{h_{i\chi}^{(rf)}}{\omega^{(k)} - \omega}$$

$$C_k^{(-\omega)} = -i\gamma V_{i\alpha}^{(k)*}\varepsilon_{\alpha\beta\chi}m_{i\beta}^{(0)}\frac{h_{i\chi}^{(rf)*}}{\omega^{(k)} + \omega}$$
(3.8)

Keeping in mind that we always have complex conjugate eigenvectors, we label those associated with an eigenfrequency having a positive real part with positive k, and those associated with a negative real part with negative k; we may then write

$$C_{-k}^{(\omega)} = C_k^{(-\omega)*}$$
$$V_{i\alpha}^{(-k)*} = V_{i\alpha}^{(k)}$$
$$\omega^{(-k)} = -\omega^{(k)*}$$
(3.9)

Since $C_{-k}^{(\omega)}$ is always off-resonance, and therefore is extremely small we can write

$$m_{i\alpha}^{(1)}(t) = V_{i\alpha k}\left(C_k^{(\omega)}e^{-i\omega t} + C_{-k}^{(-\omega)}e^{i\omega t}\right) \approx \underset{k>0}{\text{Re}}\left(V_{i\alpha k}C_k^{(\omega)}e^{-i\omega t}\right)$$
(3.9)

Note that the applied $\mathbf{H}_1$ field need not be uniform and may have any spatial dependence consistent with the magnetostatic conditions $\nabla \cdot \mathbf{B} = 0$ and $\nabla \times \mathbf{H} = 0$. The same holds for the static field, $\mathbf{H}_0$.

Recalling the thermodynamic expression for the change in magnetic energy,[x] we may write the absorption as:

$$\dot{E} = 2\lim_{t\to\infty}\frac{1}{t}\int \sum_i \frac{dm_i}{dt}\cdot H_i dt = \omega V \sum_i \text{Im}\left(m_i^{(1)} \cdot h_i^{(rf)*}\right)$$
(3.10)

where V is the volume of the object. This can be shown to be proportional to $\left(h^{(rf)}\right)^2$; all the quantities occurring in these formulas can be found by using the method given in previous section, which gives

$$\dot{E} = \omega V \gamma \, \text{Re}\left(-\frac{V_{l\beta}^{(k)*}\varepsilon_{\beta\sigma\chi}m_{l\sigma}^{(0)}h_{l\chi}^{(rf)}}{\omega - \omega'^{(k)} + i\beta\omega''^{(k)}}V_{i\alpha}^{(k)}H_{i\alpha}^{(rf)*}\right).$$
(3.11)

This formula can be used to calculate the absorption as a function of both the frequency of the r.f. field and the strength of the static magnetic field. Note that the quantity

$$a_k = -V^{(k)*}_{l\beta} \varepsilon_{\beta\sigma\chi} m^{(0)}_{\cdot l\sigma} h^{(rf)}_{\cdot l\chi} V^{(k)}_{i\alpha} H^{(rf)*}_{i\alpha} \qquad (3.12)$$

plays the role of an *oscillator strength* – it can be calculated separately; for a given $\omega$ $\dot{E}$ then involves only a single sum in Eq. (3.12). Absorption is determined by the imaginary part of oscillator strength, with the sign chosen in such a way that positive values of the imaginary parts of oscillator strength and positive values of $\dot{E}$ corresponds to absorbed energy.

It is easy to see that symmetric fields excite mostly symmetric modes and anti-symmetric fields excite mostly anti-symmetric modes – otherwise the sum in Eq.(3.12) vanishes. It can also be shown that the width of the absorption lines is proportional to $\beta\omega''$.

Depending on the initial configuration of the system, one can identify 3 distinct cases:
   a. All oscillator strengths correspond to positive absorbed energy. This means that the system is in equilibrium. If a small r.f. field is applied to the system, it moves to a dynamical steady state configuration for which the energy absorbed by the system equals the energy received from the field. It's possible that due to discretization errors some of the oscillator strength will be slightly negative, even when the system is initially at the equilibrium, however the value of these oscillator strengths will be many orders of magnitude smaller than the value of oscillator strengths for the most excited modes.
   b. Some of oscillator strengths have negative imaginary parts, but all the imaginary parts of the eigenvalues are negative. The system is then not in the equilibrium. Applying r.f. fields that excite the modes with negative oscillator strengths will move the system away from the initial configuration, raising the system energy in the process.
   c. Some of the imaginary parts of eigenvalues are positive. The system is intrinsically unstable. Excitation of modes with positive imaginary parts of eigenvalues leads to divergent behavior.

There has been much interest in the past decade in devices that exploit charge currents having an accompanying spin current (the broad field being now referred to as Spintronics). Here we restrict to a specific class of such devices involving the injection of a spin-polarized current into a ferromagnetic sample. To describe the resulting interaction between the magnetization and the spin current one can introduce a phenomenological term in Landau-Lifshitz equation so as to yield a semi-classical description (quantum mechanical models can be constructed which yield the adopted form[xi]). With this additional term the Landau-Lifshitz equation can be rewritten as

$$\frac{d\mathbf{m}}{dt} = -\gamma \mathbf{m} \times \mathbf{h} \quad (3.13)$$

$$\mathbf{h} = \mathbf{h}^{true} + \frac{\mathbf{m}}{M_s} \times \left( \beta \mathbf{h}^{true} - \mathbf{h}_J \right) \quad (3.14)$$

where $\mathbf{h}$ is an effective magnetic field involving: $\mathbf{h}^{true}$ (the sum of the external, exchange, and anisotropy fields); the effects due to damping; and a field $\mathbf{h}_J$ arising from the spin polarized electrical current. The interaction between the magnetization and the electrical current has here been modeled as a *spin transfer torque* [xii,xiii,xiv,xv,xvi], $\frac{\gamma}{M_s} \mathbf{m} \times (\mathbf{m} \times \mathbf{h}_J)$, where $\mathbf{h}_J = a_J \hat{\mathbf{m}}_J$, $\hat{\mathbf{m}}_J$ is a unit vector the direction of spin polarization, and $a_J$ is an empirical factor measuring the strength of the coupling (in units of magnetic field[4] where 1000 Oe corresponds to $10^8$ A/cm$^2$).

The excitation of spin waves due to a current then involves an effective $\mathbf{h}_1$ excitation field of the form

$$\mathbf{h}_1 = -\frac{\mathbf{m} \times \mathbf{h}_J}{M_s} \quad (3.15)$$

The contribution to the specific heat due to spin waves can be obtained by the standard prescription of assigning an energy $E_i = \hbar \omega_i$ to each mode with the occupation weighted by the Bose factor,

$$P(\omega_i, T) = \frac{1}{e^{\hbar\omega_i/kT} - 1}. \qquad (3.16)$$

The resulting expression for the energy is then given by the usual form

$$E(T) = \sum_{i=1}^{N} \frac{\hbar\omega_i}{e^{\hbar\omega_i/k_B T} - 1} \qquad (3.17)$$

where $k_B$ is Boltzmann's constant and N is the total number of spin waves (determined by appropriate scaling of the number of dipoles used in the calculation); the specific heat is then

$$C(T) = \frac{\partial E}{\partial T} = \sum_{i=1}^{N} k_B \left(\frac{\hbar\omega_i}{k_B T}\right)^2 \frac{e^{\hbar\omega_i/k_B T}}{\left(e^{\hbar\omega_i/k_B T} - 1\right)^2}. \qquad (3.18)$$

It may be of interest to examine deviations from the bulk specific heat that arise in arrays of nano-particles using this computational approach (which intrinsically includes surface effects).

**4. Application to magnetic nanodots.**

We have applied the method described here to calculate the absorption spectra and resonant modes in four representative magnetic nanodots: a disk, D = 175 nm outer diameter, L = 25 nm thickness; a ring, D = 175 nm outer diameter, inner diameter 25 nm, L = 25 nm thickness and a square slab 150×150×25 nm. In all cases we chose permalloy as the material, which has the following parameters: saturation magnetization $M_s = 795\,\text{emu/cm}^3$, exchange stiffness $A = 1.3 \cdot 10^{-6}$ erg/cm, and damping coefficient $\gamma = 0.01$. For individual cells we chose 2.5×2.5×2.5 nm blocks, followed by averaging over the sample thickness, which corresponds to studying only the modes that are uniform in the direction perpendicular to the plane of the disc.

Consider the case where no external d.c. magnetic field is present; the equilibrium configuration in this case is then a so-called vortex state, which has been described previously in some detail[xvii,xviii,xix]. It is characterized by a localized structure (the vortex core) in the center of a circular or square slab where the spins mostly point out of the plane of magnetization, whereas the remainder of the spins lie on closed loops around the core. In rings the core is absent, with the rest of the sample having a distribution of magnetization similar to that in disks. A plot of the absorption for the above mentioned samples in a uniform r.f. field applied in the sample plane is shown in figure 1.

Disks and squares exhibit an extremely low frequency mode, well separated from the rest of the spectrum. This mode corresponds to an oscillation of the vortex core, and is traditionally called a "Gyroscopic mode". The properties of this mode have been studied theoretically, experimentally and numerically by many authors[xx,xxi,xxii,xxiii,xxiv,xxv].

We calculated the frequency of the gyroscopic mode as a function of the disk diameter, with the diameter to thickness ratio held constant at D/L = 10. The calculations resulted in the frequency of the gyroscopic mode decreasing slightly with the increasing disk diameter; for given parameters it is approximately equal to 0.8 GHz. To examine the effect of discretization errors, we increased the number of dipoles by the factor of 2 with the result that the frequency changed by only 0.2%.

The behavior we obtain for the frequency of the gyroscopic mode is to some extent consistent with the experimental results of Park[xvii] who reported it to be in the range of 0.6 GHz, with the difference possibly attributed to our choice of the D/L ratio, and with the theoretical predictions of Guslienko[xx]. However it is inconsistent with calculations by Hertel[xviii], where a much lower value, 0.2 GHz was given. It is also quite different from some theoretical treatments, which predict a rapid decrease in the frequency as D increases; for example according to Ivanov[xiv]:

$$f = \frac{1}{2\pi} 4\pi\gamma M_s \left(\frac{2l_0}{D}\right)^2 \left(\frac{\Lambda-1}{\Lambda+1}\right)$$
$$\Lambda = \frac{DL}{4\pi l_0^2} \ln\left(\frac{2D}{L}\right)$$
(4.1.)

where $l_0$ is effective exchange length – about 4.8 nm for permalloy. As can be seen, this formula gives values for the frequency that are a few times (around 0.1 GHz vs. 0.8 GHz) lower than those resulting from our calculations.

The remainder of the modes for the disk can be approximately described as:

$$m^{(1)} = e^{im\varphi} F_n(r) \qquad (4.2)$$

where $\varphi$ is in-plane angle, m and n are integer numbers, and $F_n(r)$ is a function of the distance from the disk's center[xiv, xvi]. In our case we achieved good agreement with the theory, in particular concerning the spatial distribution of the oscillations, the dependence of the frequencies on the disk diameter. There are two important exceptions: Firstly, the oscillations do not disappear at the boundaries (a similar result was reported by Giovannini [xix]). Secondly, some of the modes with odd symmetry do not form doublets (theory predicts that the frequency depends only slightly on the sign of angular mode number m, the only dependence being due to the existence of the vortex core). This is due to the fact that we used square lattice in our numerical simulations, and therefore the modes with different orientation with respect to the lattice' axis have different frequencies. It should be noted that nearly all of the modes excited by the in-plane r.f. field have non-zero amplitudes in the vortex core – this is because in the vortex core the oscillations are confined to the disk's plane and therefore can very effectively couple to the in-plane r.f. field. Also, the modes with even m appear as standing waves, with $e^{im\varphi}$ in Eq. (4.2) replaced by cosine and sine functions, but modes with odd m appear as running waves.

In figures (2.I-III) we show the most strongly excited mode for the disk; the images are completely consistent with experimentally obtained images reported by Zhu et al[xxvi]. Here and in the following graphs, only z projection of the modes is being depicted, with colors corresponding to the magnitude of oscillations. Note that opposite sides of the sample are magnetized in opposite directions, characteristic of a vortex. Therefore if, for a given mode, opposing sides have opposite phases, then when projected onto the same coordinate xis the oscillations would actually be in phase. In our case, the excited modes can be classified as: 2.I. – gyroscopic mode (m=1, n=0); 2.II. m=1, n=1 mode; 2.III m=1, n=2 mode.

When we discuss the modes in the ring, the major difference from the disk is the absence of the vortex core. All the doublets become degenerate (as we mentioned before, the use of square lattice prevents some of the doublets from being completely degenerate). At the same time the number of modes that can be excited by in-plane r.f. fields greatly decreases. In disks, modes can be excited if their spatial distribution obeys certain symmetry requirements; since the precession occurs in local coordinate systems, the mode should possess odd symmetry in order to be excited, Alternatively the mode can be excited if much of its amplitude is concentrated in the vortex core, since the magnetization inside the vortex core is mostly perpendicular to the plane of the disc, and therefore it couples to the in-plane r.f. field. Only modes of the first type can be excited in the case of rings; therefore the spectrum is much narrower and corresponds to the excitation of fewer modes. In Figures 2.IV. and 2.V. we show the most strongly excited doublet for the ring.

Spin waves in square slabs have previously been studied experimentally by Perzlmaier et al[xxvii]. The main difference with respect to the absorption spectrum for disks arises from the presence of low-frequency "corner" modes (figures 2.VII and 2.VIII), i.e. modes that are confined mostly to the corners. The physical meaning of these modes is that these are the modes localized along the domain walls (vortex state in square slabs can also be represented as a state formed by four domain state). These modes have been recently observed by Demokritov. If the vortex core is removed (e.g. a square nanodot with a hole), these corner modes are doubly degenerate – one with odd and one with even symmetry. Other modes are affected by the shape changes to a lesser degree, especially those modes with 4-fold symmetry, which remain nearly exactly the same as for disks. In figures 2.VI-X. we show the most strongly excited mode for the square slab.

**5. Conclusions.**

We have developed a method for calculating the resonant modes and the absorption characteristics of a magnetic body of arbitrary shape that is applicable in the linear regime. The techniques have been demonstrated for vortex structures in discs, rings and square nanodots.

The program that incorporates our code is available for public use at www.rkmag.com. We thank G. Finocchio, P. Sievert and K. Guslienko for discussions.

**Acknowledgments.**

This work was supported by the National Science Foundation under grant ESC-02-24210.

# Figure Captions

**Figure 1. Absorption spectrum: disk (D = 175nm, L= 25nm), ring (D = 175 nm, inner diameter 35nm, L = 25nm), square nanodot (150×150×25nm) the r.f. field is uniform and is in-plane (along a principal axis for the squares).**

**Figure 2. Z projection of most strongly excited modes for: disk, (I) f=1.05 GHz, (II) f=10.11 GHZ, (III) f=12.34 GHz; ring, (IV) and (V) f=10.5 GHz; square slab, (VI) f=1.04 GHz, (VII) f=6.27 GHz, (VIII) f=10.34 GHz, (IX) f=12.22 GHz, (X) f=13.05 GHz.**

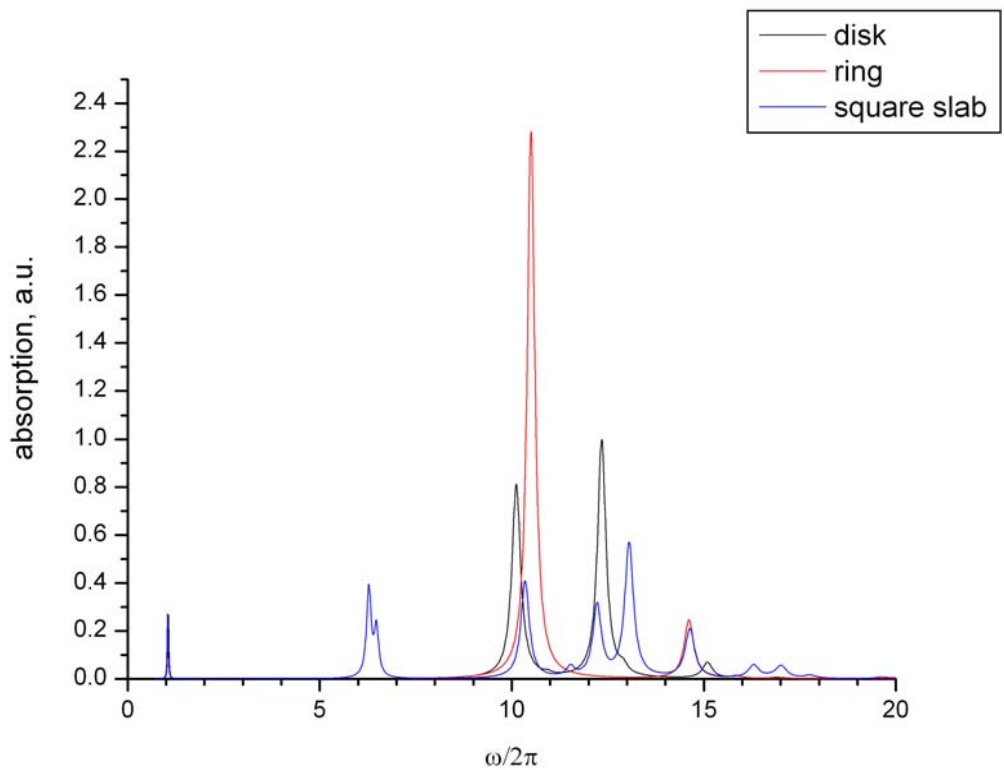

**Figure 1. Absorption spectrum: disk (D = 175nm, L= 25nm), ring (D = 175 nm, inner diameter 35nm, L = 25nm), square nanodot (150×150×25nm) the r.f. field is uniform and is in-plane (along a principal axis for the squares).**

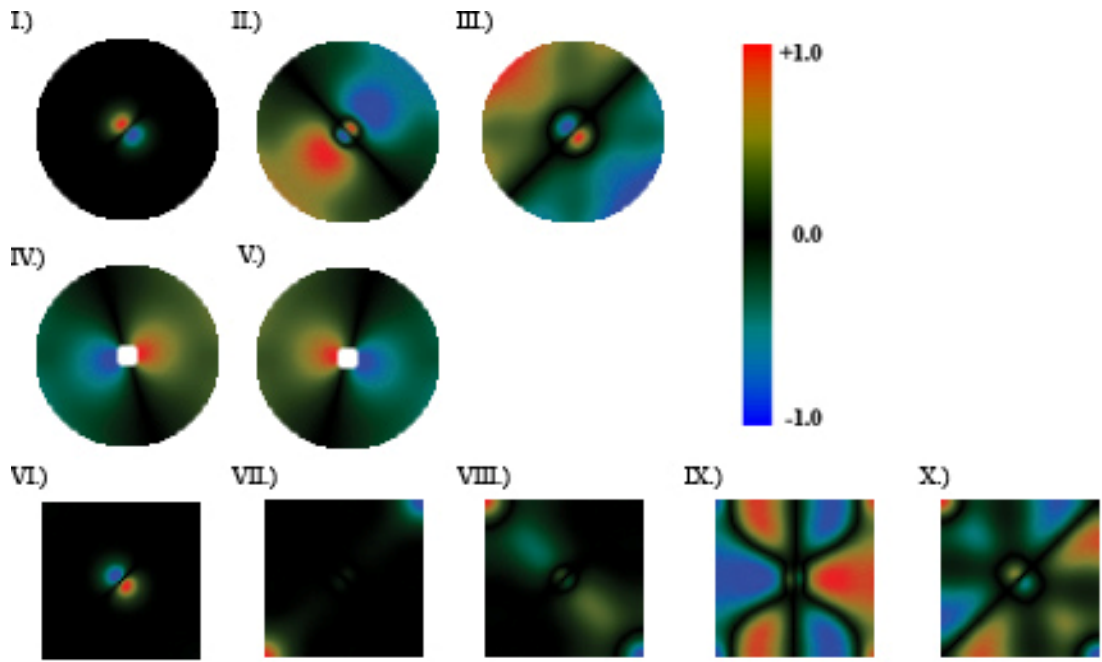

Figure 2. Z projection of most strongly excited modes for: disk, (I) f=1.05 GHz, (II) f=10.11 GHZ, (III) f=12.34 GHz; ring, (IV) and (V) f=10.5 GHz; square slab, (VI) f=1.04 GHz, (VII) f=6.27 GHz, (VIII) f=10.34 GHz, (IX) f=12.22 GHz, (X) f=13.05 GHz.